\input amstex   
\documentstyle{amsppt}   
\magnification 1200
\topmatter
\title
Wave Particle Duality in General Relativity.
\endtitle  
\author Paul O'Hara  
\endauthor  
\affil Northeastern Illinois University\\  
5500 North St. Louis Avenue\\  
Chicago, IL 60625-4699  
\endaffil 
\endtopmatter
\document
\baselineskip 1em
\TagsOnRight   
\flushpar {\bf Abstract:} In this paper a one to one
correspondence is
established between space-time metrics of general relativity and
the wave equations of quantum mechanics. This
is
done by first taking the square root of the metric associated
with
a space
and from there, passing directly to a corresponding
expression in the dual space.  It is shown that in the case of a
massless particle, Maxwell's equation for a photon follows while
in
the case of a particle with mass, Dirac's equation results as a
first
approximation. Moreover, this one to one correspondence suggests
a natural explanation of wave-particle duality.  As a
consequence, the distinction between quantum mechanics and
classical
relativistic mechanics is more clearly understood and the key
role
of initial conditions is emphasized.\newline
PACS NUMBERS: 03.65, 04.60
\newline\newline
\flushpar {\bf I Introduction}

In this paper a one to one correspondence is established between
space-time metrics of general relativity and the wave equations
of
quantum mechanics. This is done by first
taking
the square root of the metric associated with a space.  In this
way, spin is introduced in a natural way into the space-time
metric
and would seem to be equivalent to the approach of
Cordero, Tabensky and
Teitelboim [1, 2, 3] in their formulation of a theory of
supergravity. They introduce the notion of spin ``into general
relativity by taking the square root, a l$\grave a$ Dirac, of the
Hamiltonian constraints of the theory''[4]. The paper differs in
that we take the square roots of metrics (not Hamiltonians) and
focus primarily on the relationship between metrics and particle-
wave equations. We also emphasize the probability aspects of the
problem and the key role of initial conditions. As a
consequence, the distinction between quantum mechanics and
classical relativistic mechanics is more clearly understood.

Once the square root of the metric is taken, it is easy to pass
directly to a corresponding
expression in the dual space. This ``corresponding expression''
in
the dual space will oftentimes be referred to as a `` wave
equation''. It is shown that in the case of a massless particle,
the wave equation for a photon follows, while in the case of a
particle with mass, Dirac's equation results as a first
approximation.  Finally, we focus on the fact that the factoring
technique which gives rise to the square root of the metric is
not
unique and allows us an alternative way of interpreting the
negative energy levels associated with the solutions of the Dirac
equation.
\newline\newline \flushpar {\bf II Gravity and Spin}

Goudsmit and Uhlenbeck were the first to introduce the notion of
spin into quantum physics.  Their work was eventually developed
further by Pauli who was able to formulate a matrix
representation of the spin operator.  Later with the introduction
of the Dirac equation it was found that spin was a relativistic
effect obtained by linearizing the Hamiltonian of special
relativity.  The corresponding spin matrices which
resulted from this linearization also helped to explain the
peculiar electric and magnetic moments associated with the motion
of an electron in an electromagnetic field. However, in the light
of the geometrical and isotropical nature of spin [5], it also
suggests that spin is intrinsically linked to the
geometrical properties of space-time and as a consequence
gravity. This is even more evident when we
linearize not the Hamiltonian of relativity theory as did Dirac,
but rather the space-time metric itself. In fact, given the usual
metric of Minkowski's space
$$ds^2=dx^2_0-dx^2_1-dx^2_2-dx^2_3\tag2-1$$
where $dx_0=cdt$ and linearizing it in the same way as the
Hamiltonian, we find that
$$ds =\alpha_0 dx_0 +i\alpha_1 dx_1 +i \alpha_2 dx_2 +i \alpha_3
dx_3.\tag2-2$$
On squaring this out and equating it to 1-1, gives for each
$\mu,\quad \nu \in \{0,1,2,3,\}$
$$\alpha^2_{\mu}=1\qquad \text{and}\qquad \alpha_{\mu}
\alpha_{\nu}+\alpha_{\nu} \alpha_{\mu} =0.$$
These operators, therefore, are identical to the operators that
are
obtained from the Dirac equation and so spin can be defined in
the
usual way by putting
$$\sigma_1 =-i\alpha_2 \alpha_3,\qquad \sigma_2=-i\alpha_3
\alpha_1 \qquad  \sigma_3=-i\alpha_1 \alpha_2.$$
This relationship between the metric structure and the special
relativistic Hamiltonian should come as no surprise. Both the
metric and the Hamiltonian are covariant under the Lorentz
transformation and as such reflect the same geometrical
properties. However, if this is to be true in general, then we
should expect the wave equations associated with a given space to
reflect the underlying geometrical structure. In other words the
space-time structure should be an effective cause [6] of the wave
equation, in the sense that once the wave equation is given the
metric can be immediately written down and once the metric is
given
the wave equation can be written down. Moreover, the
resulting wave equation will be relativistically covariant by
definition.  The objective of the next section is to write down
these globally covariant wave equations.  The essential idea lies
in the fact that if $h dx$, where $h$ is a constant associated
with
a curvilinear coordinate, lies in some dual vector space then
this
can be associated in a unique way with the differential operator
$\partial/ h\partial x$ in the original space. For example, if
the
euclidean metric in curvilinear coordinates is of the form   
$$\vec ds = h_1 dx_1\vec e_1 + h_2 dx_2\vec e_2 + h_3 dx_3\vec
e_3$$
where $h_1, h_2, h_3$ are the curvilinear coefficients, $\vec
e_1$,
$\vec e_2$, $\vec e_3$ are unit vectors, then the corresponding
``particle-wave'' equation can be
expressed as
$$ \bigtriangledown \psi=\frac {1}{h_1 }\frac {\partial
\psi}{\partial x_1}\vec e_1+\frac {1}{h_2 }\frac {\partial
\psi}{\partial
x_2}\vec e_2+\alpha_3\frac
{1}{h_3}\frac {\partial \psi}{\partial x_3}\vec e_3,$$ with the
actual
form of the $\bigtriangledown $ operator being determined from
the
physics of the situation (see below).  Note, also, that squaring
out this latter equation gives the Laplace operator. 

In the next section,
we establish the formal relationship between the metric and the
``wave'' equation, for the general relativistic case. We then
proceed in subsequent sections to explore this rapport for some
concrete examples.\newline\newline
\flushpar{III \bf From Particles to Waves}
We shall take as our starting point the natural canonical
correspondence that exists between the differential 1-forms of
the
type $dx$, used in defining the metric, and the covariant tensors
over a manifold of the type  $\frac
{\partial}{\partial x}$. It is precisely this canonical 1-1
correspondence that allows us to define the wave equations we
seek.

Once the wave equations are obtained we will then show that in
the
special relativistic case this gives rise to the wave equation
for
spin 0 massless particles and to the usual Dirac equation for
neutrinos. In the case of particles with mass, however, both the
Klein-Gordan and the general Dirac
equation will be seen to be only first approximations of the more
general wave equations.\newline
Let $C$ be the set of complex numbers and $\Cal F (M)$ be the set
of twice differentiable functions over a manifold $M$. Define a
vector field $v$ as a map 
$$v: \Cal F (M)\to \Cal F (M) :f  \mapsto vf. $$ with properties
$$
v (af + bg) = avf + bvg \tag 3-1 $$
$$ v(fg)=f(vg)+g( vf) \tag 3-2$$
where $f\in \Cal F (M)$, $g\in \Cal F (M)$ and $a,\ b \in C$.
Denote the set of vector fields $\{v \}$ by $T^1_0(M)$.  For the
purpose of this paper we will work with a 4-dimensional pseudo-
Riemannian manifold, with $x_0=ct$ where $t$ is the local time
and
$c$ is the velocity of light.  We will also use Einsteinian
notation throughout for the indices; in other words,
$a^{\mu}b_{\mu }=\sum_{\mu }a^{\mu }b_{\mu }$, $\mu \in \{0,1,2,3
\}$. This means that we can represent $v\in T^1_0(M)$ in terms of
a local coordinate system as $v=v^{\mu }\frac
{\partial}{\partial x^{\mu }}$ where $v^{\mu } \in \Cal F (M)$ .
We
likewise, define the differential one forms as a map
$$\omega : T^1_0 \to \Cal F (M): v \mapsto \omega (v)$$ 
with the properties
$$ \omega (u+v) = \omega (u) + \omega (v)$$
$$ \omega (fu) = f \omega (u) \qquad f \in \Cal (M).$$
We denote the space of 1-forms by $T^0_1$. In terms of a local
coordinate system, if $df \in T^0_1$ then we can write this as
$df
= f_{,\mu}dx^{\mu } = \frac {\partial f}{\partial x^{\mu }
}dx^{\mu
}.$  

Given a metric tensor we can define a 1-1 canonical
correspondence between elements $T^1_0 (M)$ and $T^0_1 (M)$ by
the
map
$$ \wedge : T^1_0 \to T^0_1 : u \mapsto \overset \wedge \to u $$
where $\overset \wedge \to u(v) = <u,v>,\quad \forall v \in T^1_0
(M).$ In particular, if $\overset \wedge \to u = \overset \wedge
\to u_{\mu }dx^{\mu }$ then we can identify $\overset \wedge \to
u$
and  $u$ and write $u_{\mu }=g_{\mu \nu}u^{\nu }$. Similarly, if
$\vee $ is the inverse of $\wedge $, i.e.
$$ \vee : T^0_1 \to T^1_0 : \omega \mapsto \overset \vee \to
\omega
$$ where $<\overset \vee \to \omega,v>=\omega (v),\quad \forall v
\in T^1_0 (M).$ In particular, if $\overset \vee \to \omega
=\overset \vee \to \omega_{\mu }\partial^{\mu }$ then we can
identify $\overset \vee \to \omega$ and  $\omega$ and write
$\omega^{\mu }=g^{\mu \nu}\omega_{\nu }$.

By means of these canonical transformations we can now easily
pass
from particles to waves and vice-versa, or more precisely we can
pass from metrics associated with the particles to the
corresponding wave equation for the particle.  In general, the
form
of the metric associated with a particle is
$$ds^2=g_{\mu \nu}dx^{\mu } dx^{\nu }. \tag3-3$$
Since $g_{\mu \nu}$ is a symmetric matrix, its square root
exists. The square root is symmetric but not unique and the
significance of this non uniqueness is discussed in the last
section of this paper. For the moment we denote this square root
matrix by
$h_{\mu \nu}$ and note that $h_{\mu \nu}=h_{\nu \mu}$. 

The linearized metric can now be written as a spinor: $ds = h_{\mu
\nu}\alpha^{\mu} dx^{\nu }$, where $(\alpha^{\mu})^2=1$ and
$\alpha^{\mu} \alpha^{\nu}+\alpha^{\nu}\alpha^{\mu}=0$. Note also
that if $\gamma_{\mu}=h_{\mu \nu}\alpha^{\nu}$ then $\gamma_{\mu}
\gamma_{\nu} + \gamma_{\nu}\gamma_{\mu} = 2g_{\mu \nu}$. 
It follows by the
canonical correspondence that the
associated particle wave equation will be given by:
$$\frac {\partial \psi}{\partial s}=h^{\mu \nu}\alpha_{\mu
}\frac {\partial \psi}{\partial x_{\nu} }.\tag3-4$$
Note that $\alpha_{\mu } =\alpha ^{\mu }$ for each $\mu $ and
that
$ds$ and hence $\frac {\partial }{\partial s}$ are covariant
under
coordinate transformations, according to the rules for a tetrad
formalism.[7]  Equation 3-4 will be seen as the most general form
of a particle-wave equation.  In effect, it describes the motion
of
a particle from a reference frame within the field of the
particle.
The wave function $\psi$ could represent a function describing
the
distribution of matter within a classical object, such as a star,
as seen from the reference frame. Similarly, if the initial
conditions are unknown, then $\psi$ could represent the
probability
distribution associated with the initial position of the object.
This is the case with
elementary particles where the initial positions, on account of
the
uncertainty principle, are in principle unknowable. Finally we
note
that if the rest mass $m$ of the particle is a constant, $h$ is
Planck's constant and $i=\sqrt -1$, then if we seek
solutions of the form $\frac {\partial \psi}{\partial s}=k\psi $
where $k=2\pi \frac {imc}{h}$,
equation 3-4 can be reduced to
$$mc^2\psi =-i\frac {h c}{2\pi }\alpha_{\nu }
h^{\mu \nu}\frac {\partial \psi}{\partial x_{\nu}}. \tag3-5$$ We
will now use equation 3-4 (3-5)to investigate specific types of
equations for specific types of metrics.\newline \newline
\flushpar {IV \bf Photon and Neutrino Equations}

The linearized metric for a massless particle is given by   $$ 0=
\alpha^0 cdt-\alpha^1dx_1-\alpha^2dx_2 -\alpha^3dx_3$$ from which
it follows by the canonical correspondence established above that
the associated wave equation for the particle is given by:
$$0 =\alpha_0\frac {\partial \psi}{c\partial t}-\alpha_1\frac
{\partial \psi}{\partial x_1}-\alpha_2\frac {\partial
\psi}{\partial x_2}-\alpha_3\frac {\partial \psi}{\partial
x_3}.$$
This is the Dirac equation for a massless particle. 
Squaring this out we get Maxwells equation (or the Klein-Gordan
equation for a massless particle) namely:
$$ \frac {1}{c^2}\frac {\partial^2 \psi }{\partial
t^2}=\sum_{i=1}^3 \frac {\partial^2 \psi}{\partial x^2_i}.$$ Note
that, in this formulation, solutions of the massless Dirac
equation
are also solutions of the usual massless Klein-Gordan equation
(Maxwell's equation).
\footnote {It has been noted in a previous paper [8]
that Fermi-Dirac statistics is a consequence of particle coupling
and Bose-Einstein statistics is a consequence of decoupled
particles.  Moreover, it follows as a trivial consequence of that
result that bosons cannot be second quantized as fermions and
fermions cannot be second quantized as bosons.  In other words,
particles which at time t are coupled with probability one cannot
at the same time, t, be coupled with probability less than 1
(decoupled). They are either in one state or another. However, it
is possible to make and break couplings. Transposed into the
context of quantum field theory this means that the wave function
of coupled particles will have the anti-commutator =0 (the
singlet state being a case in point) while the wave fuction of
the
decoupled particles will have the commutator =0. [9]}

Since the wave equation emerges from the structure of space-time
itself, the question arises as to how to distinguish classical
mechanics from quantum mechanics. We investigate this by
analyzing
the motion of a photon in a Minkowski space,
subject to different sets of boundary conditions. In the first
case
we consider the motion of a photon moving on the x-axis with
uniform velocity $c$, but constrained by two mirrors placed at
$x=0$ and $x=\xi$ to move uniformly on the interval $[0, \xi]$.
We
will assume that perfect reflection takes place at the mirrors
and
that no energy is exchanged.  In this case, if the photon were a
strictly classical particle with position $x=0$ at $t=0$ then its
equation of motion would be of the form:
$$x=\cases ct - 2n\xi, &\text {for $t\in [\frac {2n\xi}{c}, \frac
{(2n+1)\xi}{c}]$}\\
2(n+1)\xi -ct &\text {for $t\in [\frac {(2n+1)\xi}{c}, \frac
{(2n+2)\xi}{c}]$}
\endcases$$
and its wave function $\psi (x,t)$ would be of the form
$$\psi(x,t)=\cases \delta [k(x-ct)] &\text {for $\qquad x-
ct=-2n\xi$}\\ 
\delta[k(x+ct)] &\text {for $\qquad x+ct=2(n+1)\xi$}\\
0 &\text {otherwise.}
\endcases$$
The wave function in this case pinpoints the position of the
particle with probability 1. Moreover, there is no restriction on
the energy of the photon in this case. Theoretically, it may have
values ranging from 0 to $\infty$.

However, the classical particle is an idealized situation.  In
reality, the position of the photon constrained to move on the
line
is unknown and any attempt to know its exact position will be
subject to Heisenberg's uncertainty relations.  In other words,
its
exact position can not be known in principle, because any attempt
to pinpoint it will scuttle the position and defeat the whole
purpose of the experiment.  The best we can do is to
describe the position by means of a uniform probability density
$f(x-ct)=1/\xi$ for $x \in [0, \xi]$ which means $\psi (x,t)=
e^{\pm ik(x-ct)}/\sqrt \xi$.  This does not mean that causality is
violated nor that the particle  does not have an exact position.
It
simply affirms that our initial conditions have to be defined
statistically and as a consequence the future evolution of the
system is best interpretated in a statistical way. Finally, note
that in this model the energy of the particle can once again vary
from 0 to $\infty$ in a continuous manner.

Thirdly, the particle may be constrained to move in a potential
well in such a way that the wave function is continuous (= 0) at
the boundaries. In the case of the above problem, this would mean
that the wave function would have stationary solutions of the
form
$\psi(x)=c\sin \bigl(\frac {n \pi}{\xi }x \bigr)$ where $c$ is a
constant and the photon energy would be quantized and of the form
$E=h\nu $. 

The purpose of the above three examples is to highlight the
importance of the boundary conditions when distinguishing between
a classical type problem and a quantum mechanical problem, a
point
also stressed by Lindsey and Margenau [10].   
Classical and quantum laws are not in opposition to each other.
There is not one set of laws on the microscopic level and another
on the macroscopic.  On the contrary, classical and statistical
methodologies are complimentary to each other and are in
principle, applicable at all levels. However, on the microscopic
level, statistical fluctuations will be more  pronounced because
of
the uncertainty principle and in this case, the effects
associated with quantum physics will become more
apparent.\newline\newline
\flushpar {V \bf Particle Wave Equations for Particles with Mass}

We now turn our attention to particles with mass.  Our approach
will contrast with classical relativistic quantum theory in that
we
are no longer dealing with a particle existing within a given
space-time manifold (for example an electron in a Minkowski
space)
rather our space-time structure is inherently related to the
presence of the mass contained within it.  In other words, the
presence of the mass is an effective cause of the curvature of
the
space time [11] and not just an incidental presence within the
space time manifold. In this regard, the Minkowski metric of
special relativity
$$ds^2=dx_0^2-dx_1^2-dx_2^3-dx_3^2$$
cannot be an appropriate metric for any particle with mass.
Moreover, the Dirac equation which comes from linearizing the
Hamiltonian of special relativity, cannot be the proper wave
equation for any elementary particle with mass.  This follows
from
the fact that such a particle by definition curves space-time and
hence cannot be imbedded in the flat space-time which underlies
the
Dirac equation. We begin our analysis with massive particles
without charge. \newline\newline
To determine the wave equation of an neutral particle such as a
neutron, it is assumed that a free neutron is spherical. Hence,
the
metric associated with it will be the usual Schwarzschild one:
$$ds^2=\biggl (1-\frac{2Gm}{c^2r}\biggr )dx^2_0-\biggl (1-
\frac{2mG}{c^2r}\biggr )^{-1}dr^2-
r^2(d{\theta}^2+\sin^2 \theta d{\phi}^2).$$
Linearizing this we get
$$ds=\alpha_0 \biggl (1-\frac{2Gm}{c^2r}\biggr )^{\frac
12}dx_0+i\alpha_1\biggl (1-
\frac{2Gm}{c^2r}\biggr )^{- \frac
12}dr+ir(\alpha_2d{\theta}+\alpha_3 \sin {\theta}
d{\phi}).\tag5-1$$
It now follows from the canonical correspondence discussed above
that the generalized Dirac equation for the Schwarzschild metric
is
given by $$\frac {\partial \psi}{\partial s}=\alpha_0  \biggl (1-
\frac{2Gm}{c^2r} \biggr )^{-\frac 12}\frac {\partial
\psi}{\partial x_0}-i\alpha_1 \biggl (1-
\frac{2Gm}{c^2r}\biggr )^{\frac 12}\frac {\partial \psi}{\partial
r} -i\frac 1r \biggl (\alpha_2 \frac {\partial \psi}{\partial
\theta}+\alpha_3 \frac {1}{\sin {\theta}} \frac {\partial
\psi}{\partial \phi}\biggr ). \tag5-2$$
Note that no distinction has been made between a general-
relativistic-classical-type problem and a quantum-
mechanical problem. For example, solutions to the above equation
could be obtained by
analyzing the reflection of a massless particle, confined to move
on a straight line segment with endpoints $(r=r_0, \theta =0,
\phi
=0)$ and $(r=r_1, \theta =0, \phi =0)$, within the
gravitational field of the particle, in a way analogous to the
motion of a photon discussed in the previous section. However, we
will not pursue this discussion here. Instead, we will focus on
the
conditions necessary to reduce the first
approximation of equation 5-2 to the Dirac equation. This can be
done by seeking eigenvalue solutions to the equation. We will
refer
to them as equilibrium solutions. For example, in the case of the
photon problem discussed above, these solutions occurred when
continuity of the wave function was required.
 
Returning to the problem of the Dirac equation, note that the set
of equilibrium solutions $\{k \}$ are given by the equation
$\frac
{\partial
\psi}{\partial s}=k\psi$. In other words $k$ represents the set
of
eigenvalues of the system. If we now put $k=-2\pi \frac {i
mc}{h}$,
multiply the equation 5-2 by $-\frac {hci}{2\pi
}\alpha_0$ and  denote $-i\alpha_0
\alpha_i$ by $\alpha^{\prime}_i$ then for  $2m/r < 1$, we obtain
as
a first order approximation for equation 5-2: 
$$\frac {ih}{2\pi}\frac {\partial \psi}{\partial t}=-\frac
{ihc}{2\pi }
\alpha^{\prime}_1\frac {\partial \psi}{\partial
r}-\frac {ihc}{2\pi}\frac {1}{r}\biggl (\alpha^{\prime}_2\frac
{\partial
\psi}{\partial \theta} + \alpha^{\prime}_3 \frac {1}{\sin \theta}
\frac {\partial \psi}{\partial \phi}\biggr )+\alpha_0 mc^2\psi
.$$
It is easy to check that the set of operators $\alpha_0 ,
\alpha^{\prime}_i,\ i=1, 2, 3$ obey the usual algebra of Dirac
matrices. If we now take stationary solutions for the operator
$\frac {ih}{2\pi}\frac {\partial \psi}{\partial t}= E\psi$ then
Dirac's
equation follows.
At this stage a couple of interesting observations arise based on
the above analysis.  It would appear that the Dirac equation is
just an approximation of a 4-dimensional gradient times the
$\alpha
$ matrix.  From this perspective the Dirac equation could also be
interpreted in a classical or non-quantum-mechanical way.  For
example, it could represent the movement of a large planetary
size
free mass with spin. How then do we distinguish quantum mechanics
from a classical general relativistic theory?  It is comparable
to
the theory of the photon already discussed above.  The difference
rests in assigning a probability interpretation to the wave
function [12],  making full use of the
uncertainty principle, and recognizing the fundamental role of
Planck's constant as a unit of measurement in physics. It is the
choice of $h$ as a non-zero constant that causes the quantization
procedure to come about. In the photon problem discussed earlier,
the imposition of continuity on the wave function forced $h$ to
be
a non-zero constant.

A second question that arises is in the interpretation of
equation 5-2 from the perspective of general relativity; to quote
Marie Antoinette Tonnelat in this regard[13]: ``in quantum
mechanics, curved space remains a permissible framework;
according to general relativity it becomes an effective cause''.
I
claim that the above approach to the wave equation resolves this
in
a natural way.

In the approach given in this paper general relativity is the
effective cause of the form of the quantum mechanical wave
equations. For example, consider as a frame of reference, a
tetrad
with origin at the particle's center of mass. Let $\psi (r)$
represent the wave function of a massless particle in the field
of
the massive particle. The
corresponding wave equation can now be written as
$$\alpha_0  \biggl (1-\frac{2Gm}{c^2r} \biggr )^{-\frac 12}\frac
{\partial \psi}{\partial x_0}-i\alpha_1 \biggl (1-
\frac{2Gm}{c^2r}\biggr )^{\frac 12}\frac {\partial \psi}{\partial
r} -i\frac 1r \biggl (\alpha_2 \frac {\partial \psi}{\partial
\theta}+\alpha_3 \sin {\theta} \frac {\partial \psi}{\partial
\phi}\biggr )=0. \tag5-3$$
From this perspective the above equation can be taken as
describing the motion of a massless fluid in a Schwarzschild
space
of the particle.  Similarly, it may describe the motion of a
probability density function for a massless particle within the
same space. The distinction between the two cases depends on the
boundary conditions being considered. 
As a final example, we choose an arbitrary point within the
space-
time containing the particle and let $(t,r,\theta ,\phi )$ be the
position of the particle with respect to this arbitrary point.
Denote its wave function by $\psi (r,t)$, with probability
density
$|\psi(r,t)|^2$. The potential positions of the free particle are
isotropic with respect to our chosen fixed point and as such are
spherically symmetrical in a Schwarzschild space. The wave
equation
once again takes the form 5-2. If solutions of the form $\frac
{\partial \psi}{\partial s}=k \psi$ are sought and the limit as
$r\to \infty$ is taken as the special relativistic limit, then
this
will reduce to the Dirac equation if $k=\pm 2\pi i\frac {mc}{h}$.
Hence, from this perspective we can take the equation:  $$\pm
\frac
{imc}{h}\psi=\alpha_0  \biggl (1-
\frac{2Gm}{c^2r} \biggr )^{-\frac 12}\frac {\partial
\psi}{\partial x_0}-i\alpha_1 \biggl (1-
\frac{2Gm}{c^2r}\biggr )^{\frac 12}\frac {\partial \psi}{\partial
r} -i\frac 1r \biggl (\alpha_2 \frac {\partial \psi}{\partial
\theta}+\alpha_3 \sin {\theta} \frac {\partial \psi}{\partial
\phi}\biggr ) \tag5-4$$
as describing the motion of a free electron in a
``Schwarzschild space''.
\newline\newline
\flushpar {VI \bf The Hydrogen Atom}
We now apply the techniques of this paper to describe the
hydrogen atom. More specifically, we describe the motion of the
electron lying within the Reissner-Nordstrom metric [14] of the
proton.  Linearizing this metric gives:
$$ds =i\alpha_1 \biggl (1-\frac {2Gm_p}{c^2r}+\frac
{Ge^2}{c^4r^2} \biggr )^{-\frac 12}dr +ir(\alpha_2 d\theta
+\alpha_3 \sin \theta d\phi )+\alpha_0\biggl (1- \frac
{2Gm_p}{c^2r}+\frac {Ge^2}{c^4r^2}\biggr )^{\frac 12}c dt.$$    
Denoting the rest masses of the electron and proton by $m$,
$m_p$,
respectively, then the rest energy of the electron
relative to the proton will be given by $mc^2-eA_0$ where
$A_0=\frac er$ and $r$ is the distance between the proton and
electron. It follows from the usual 1-1 correspondence rule (c.f.
equation 3-5) that the particle wave equation for this metric
becomes, on multiplying across by $\alpha_0$: 
$$\align
-\alpha_0\frac {2 \pi i}{h}(mc-\frac ec A_0)\psi (r,t)=&\biggl \{
\alpha^{\prime}_1 \biggl (1-\frac {2Gm_p}{c^2r} + \frac
{Ge^2}{c^4r^2}\biggr)^{\frac 12}\frac {\partial }{\partial r} +
\frac 1r \biggl (\alpha^{\prime }_2 \frac {\partial}{\partial
\theta} +
\alpha^{\prime}_3 \frac {1}{\sin \theta}\frac {\partial
}{\partial \phi }\biggr )\\   
& \qquad +\biggl (1-\frac {2Gm_p}{c^2r}+\frac
{Ge^2}{c^4r^2} \biggr )^{-\frac 12} \frac 1c \frac {\partial
}{\partial t} \biggr \}\psi (r,t) \tag6-1 \endalign $$  
where $\alpha^{\prime}_i=-i\alpha_0 \alpha_i$. Retaining only
first
order stationary terms gives: $$ \biggl \{\alpha_0 \frac {2 \pi
i}{hc}(mc^2-eA_0)
+\alpha^{\prime}_1 \frac {\partial }{\partial r}+\frac 1r \biggl
(\alpha^{\prime}_2 \frac
{\partial}{\partial \theta} + \alpha^{\prime}_3 \frac {1}{\sin
\theta }\frac {\partial }{\partial \phi} \biggr) \biggr \}\psi(r)
= \frac  {2 \pi i}{hc}E\psi (r) \tag6-2$$
which is essentially the same as the equation of Dirac for the
hydrogen atom.  The only difference is the presence of the
$\alpha_0 eA_0$ term, instead of $eA_0$.  However, as will be
pointed out in the next section, there is a lot of arbitrariness
associated with the choice of the $\alpha $ spinors. Secondly, if
we ``square'' out equation 6-2 and denote $\frac {h}{2\pi
i}\bigtriangledown$ by $p$, we obtain the equation:
$$(c^2p^2+m^2c^4+\frac {hec}{2\pi }\alpha^{\prime}\Cal
G)\psi=(E+\alpha_0eA_0)^2\psi \tag6-3$$
The only difference between this and Dirac's equation is that we
have written $\alpha_0 eA_0$ instead of $eA_0$ on the right hand
side of the equation while on the left hand side we have written
$\frac {hec}{2\pi }\vec \alpha^{\prime}.\Cal G$ instead of $\frac
{hec}{2\pi i}\vec \alpha.\Cal G$. In other words, we have
absorbed the imaginary component into the spin matrix. Moreover,
this absorbing of the $i$ into
$\alpha^{\prime}$ is more than just a handy notation.  Earlier,
we defined $\alpha^{\prime}_i$ by $\alpha^{\prime}_i=-
i\alpha_0 \alpha_i$ and pointed out that $\{\alpha^{\prime}_i \}$
has all the same properties associated with the set $\{\alpha_i
\}$. Moreover, if we had written the rest energy of the electron
as
$mc^2-\alpha_0 eA_0$ instead of $mc^2-eA_0$, then the regular
Dirac
equation would follow. It can be argued that both are valid: the
regular Dirac case could represent coupled states of an electron-
positron combination while the other case as given by equation
6-3
could represent the coupled states of a single electron. Finally,
we point out that in both cases, the
Schrodinger equation also follows as a first approximation.
\newline\newline
\flushpar {VII \bf Negative Energy Levels Reinterpreted}\newline
\newline
The negative energy solutions of the Dirac equation are usually
interpreted in terms of virtual electrons occupying negative
energy
levels.  This may be seen as a metaphor but it also
prescinds from what is really taking place.
The first thing to notice is that neither the representation of
the
square root matrix, $h^{\mu \nu}$  nor the
representation of $\alpha_0,\  \alpha_1,\ \alpha_2, \ \alpha_3 $
are unique.  We could equally work with any representation of the
form $\epsilon_0 \alpha_0, \ \epsilon_1 \alpha_1, \ \epsilon_2
\alpha_2 , \ \epsilon_3 \alpha_3$ where $\epsilon_{\mu}$ can take
on the values $\pm 1$. In particular, if a free particle with
spin
satisfies the Dirac equation in the form: 
$$\biggl \{ c \biggl [\alpha_1 \frac {h}{2\pi i}\frac {\partial
}{\partial x}+ \alpha_2 \frac {h}{2 \pi i}\frac {\partial
}{\partial y} +\alpha_3 \frac {h}{2\pi i}\frac {\partial
}{\partial z} \biggr ] + \alpha_0 mc^2 \biggr \}\psi = E\psi ,$$
then the corresponding negative energy equation for a free
particle with spin satisfies the Dirac equation in the form:
$$\biggl \{ c \biggl [\alpha_1 \frac {h}{2\pi i}\frac {\partial
}{\partial x}+ \alpha_2 \frac {h}{2 \pi i}\frac
{\partial }{\partial y} +\alpha_3 \frac {h}{2\pi i}\frac
{\partial }{\partial z} \biggr ] + \alpha_0 mc^2 \biggr \}\psi =
-
E\psi \tag7-1$$
which is the same as
$$\biggl \{ c \biggl [-\alpha_1 \frac {h}{2\pi i}\frac {\partial
}{\partial x}- \alpha_2 \frac {h}{2 \pi i}\frac {\partial
}{\partial y} -\alpha_3 \frac {h}{2\pi i}\frac {\partial
}{\partial z} \biggr ] - \alpha_0 mc^2 \biggr \}\psi =
E\psi.\tag7-2 $$
Written in this last form we can see immediately that this too is
the equation of a free particle of POSITIVE energy but of a spin
state equal and opposite to the other particle.  In particular,
if
the particles are coupled [15, 16] then they are in mutually
opposite states with respect to the spin operator, in which case
the Pauli exclusion principle applies. It follows that as long as
the coupling lasts, the particles are not free to be in the same
state.  This coupling is implicit in the solutions of the
equations. However, if the particles are not coupled then the
particles can be in either one of the spin states and are free to
switch from one to another in accordance with the usual
probability laws but only after an interaction takes place. It
follows that the Dirac equation gives the solution for a pair of
coupled particles and implicitly contains a proof of the Pauli
exclusion principle.    
\newline \newline
\flushpar{\bf VIII Conclusion} 
We have given a heuristic approach
to unifying the theory of general relativity and quantum
mechanics. As a consequence of this unification, we have
highlighted the fact that the difference between classical
mechanics and quantum mechanics rests primarily on the choice of
initial conditions. It is the imposition of probability
conditions on a sample space that leads to quantum mechanics. 
Classical
mechanics on the other hand finds its intelligibility in its
deterministic approach.  
 
Finally, we note that to the extent that our approach is
heuristic it remains relatively invariant.  In other words, any
modifications to the theory will be to the substance of either
relativity theory or quantum mechanics. However, the general
heuristic approach should remain unchanged. 
\newline\newline
\flushpar {\bf Acknowledgements:} I would like to thank Prof.
Ephraim Frishbach of Purdue University (USA), for some useful
suggestions and for having brought to
my intention the work of Stevan Weinberg on the covariant
formulation of tetrads. I am alsoindebted to Rev. D. McCrea (deceased),
University College Dublin, Ireland, for wonderful class notes on the 
relationship between differential one-forms and the partial differential
operator.

\end